\documentclass[sigconf]{acmart}
\AtBeginDocument{%
  \providecommand\BibTeX{{%
    \normalfont B\kern-0.5em{\scshape i\kern-0.25em b}\kern-0.8em\TeX}}}

\copyrightyear{2024}
\acmYear{2024}
\setcopyright{rightsretained}
\acmConference[CHI '24]{Proceedings of the CHI Conference on Human Factors in Computing Systems}{May 11--16, 2024}{Honolulu, HI, USA}
\acmBooktitle{Proceedings of the CHI Conference on Human Factors in Computing Systems (CHI '24), May 11--16, 2024, Honolulu, HI, USA}
\acmDOI{10.1145/3613904.3641982}
\acmISBN{979-8-4007-0330-0/24/05}

\begin{document}

\title{Investigating Why Clinicians Deviate from Standards of Care: Liberating Patients from Mechanical Ventilation in the ICU}


\author{Nur Yildirim}
\affiliation{%
  \institution{Carnegie Mellon University}
  \city{Pittsburgh}
  \state{PA}
  \country{USA}}
\email{yildirim@cmu.edu}

\author{Susanna Zlotnikov}
\affiliation{%
  \institution{Carnegie Mellon University}
  \city{Pittsburgh}
  \state{PA}
  \country{USA}}

\author{Aradhana Venkat}
\affiliation{%
  \institution{Carnegie Mellon University}
  \city{Pittsburgh}
  \state{PA}
  \country{USA}}

\author{Gursimran Chawla}
\affiliation{%
  \institution{Carnegie Mellon University}
  \city{Pittsburgh}
  \state{PA}
  \country{USA}}

\author{Jennifer Kim}
\affiliation{%
  \institution{Carnegie Mellon University}
  \city{Pittsburgh}
  \state{PA}
  \country{USA}}

\author{Leigh A. Bukowski}
\affiliation{%
  \institution{University of Pittsburgh}
  \city{Pittsburgh}
  \state{PA}
  \country{USA}}

\author{Jeremy M. Kahn}
\affiliation{%
  \institution{University of Pittsburgh}
  \city{Pittsburgh}
  \state{PA}
  \country{USA}}
\email{jeremykahn@pitt.edu}

\author{James McCann}
\affiliation{%
  \institution{Carnegie Mellon University}
  \city{Pittsburgh}
  \state{PA}
  \country{USA}}
\email{jmccann@cs.cmu.edu}

\author{John Zimmerman}
\affiliation{%
  \institution{Carnegie Mellon University}
  \city{Pittsburgh}
  \state{PA}
  \country{USA}}
\email{johnz@cs.cmu.edu}

\renewcommand{\shortauthors}{Yildirim, et al.}

\begin{abstract}
Clinical practice guidelines, care pathways, and protocols are designed to support evidence-based practices for clinicians; however, their adoption remains a challenge. We set out to investigate why clinicians deviate from the ``Wake Up and Breathe'' protocol, an evidence-based guideline for liberating patients from mechanical ventilation in the intensive care unit (ICU). We conducted over 40 hours of direct observations of live clinical workflows, 17 interviews with frontline care providers, and 4 co-design workshops at three different medical intensive care units. Our findings indicate that unlike prior literature suggests, disagreement with the protocol is not a substantial barrier to adoption. Instead, the uncertainty surrounding the application of the protocol for individual patients leads clinicians to deprioritize adoption in favor of tasks where they have high certainty. Reflecting on these insights, we identify opportunities for technical systems to help clinicians in effectively executing the protocol and discuss future directions for HCI research to support the integration of protocols into clinical practice in complex, team-based healthcare settings.
\end{abstract}



\maketitle


\section{Introduction}
Healthcare advances result from carefully controlled, randomized, double-blind studies. Actions and interventions that prove successful in these trials get captured as clinical guidelines, care pathways, and protocols. These structures provide evidence-based scaffolding for clinicians in support of their practice, enabling the results of clinical trials to improve health outcomes at the population level. While protocols improve patient outcomes and quality of patient care, their implementation and adoption by frontline care providers remains a persistent challenge \cite{stoneking2011sepsis, gary2021addressing}. Protocols are produced in carefully controlled environments; hence, they often do not account for the messy complexity of day-to-day clinical practice \cite{gary2021addressing}, creating barriers to adoption. Specifically, protocols often cannot account for the innumerable complexities of healthcare in the real world in which care needs to adapt to specific contexts and available resources \cite{kavanagh2016standardized}. 
Recent research shows that changes to information technology (IT) systems may offer one way to increase adherence to protocols \cite{anderson2019clinical}.

In this study, we set out to investigate the adherence to clinical protocols with a specific focus on opportunities where technology might aid protocol adoption. We used mechanical ventilation in the intensive care unit (ICU) as a model system, enabling us to integrate our complementary perspectives as human-computer interaction (HCI) researchers and critical care medicine practitioners.
Mechanical ventilators save lives by taking over the work of breathing, so a patient in the ICU can rest and heal \cite{hickey2020mechanical}.
It is important to liberate patients from the ventilator as soon as possible, while also taking care not to liberate them too soon, before their lungs have had a chance to heal. Achieving this balance is a complicated and high-risk endeavor, requiring the careful consideration and coordinated actions of a team of clinicians, including a critical care physician, an ICU nurse, and a respiratory therapist (RT).

To aid this process, medical researchers created \textit{the Wake Up and Breathe} (WUB) protocol, an evidence-based guideline for assessing mechanically ventilated patients and deciding when they should be liberated \cite{khan2014effectiveness}. The protocol involves two sequential, interdependent actions. First, the ICU nurse assigned to a specific patient performs a ``Spontaneous Awakening Trial'' (SAT). They cut off a patient's sedation to observe if they can tolerate being awake. Next, while the patient is free from sedating medications, an RT performs a ``Spontaneous Breathing Trial'' (SBT). They cut off the ventilator's breathing support to observe if the patient can successfully breathe on their own without mechanical support. Based on the results of these two assessments, the physician, RT, and nurse collectively decide whether a patient should be extubated (have a breathing tube removed), liberating them from the ventilator \cite{willdenbest}.

Clinical trials show that application of WUB dramatically improves patient outcomes, reducing the average number of days a patient remains on a ventilator and decreasing mortality \cite{khan2014effectiveness, girard2008efficacy}. Unfortunately, the protocol is often not followed in clinical practice.
We chose to conduct a field study to investigate the current workflows and practices for mechanically ventilated patients. Our goal was twofold:

\begin{enumerate}
    \item To gain a deeper understanding of the barriers to WUB adherence; to help explain how, when, and why clinicians deviate.
    \item To identify opportunities where technology might effectively remove barriers and increase adherence to standards of care.
\end{enumerate}

We conducted observations and interviewed clinicians who work at three different medical ICUs within a large, integrated health system. Our findings indicate that unlike what prior literature suggests, clinicians do not seem to consciously avoid executing WUB in relation to specific barriers. Instead, the uncertainty surrounding individual patients' eligibility and the role of individual providers within the context of the interprofessional care team seems to make clinicians `procrastinate'. They unconsciously deprioritize WUB by focusing on other tasks where they have high certainty, and this causes them to miss the window for performing WUB. Specific sources of uncertainty included which patients might be eligible, when an RT might arrive to perform SBT, if the physician wants WUB to happen, and what counts as a sedative. Reflecting on these insights, we identified several opportunities where new technical systems might help. In addition, we observe that the challenge of uncertainty and adherence appears to be a problem that goes beyond liberation from mechanical ventilation and beyond clinical practice in the ICU.

This study makes two novel contributions to the HCI literature. First, our field study provides a rare, first-hand description of nurses' and RTs' clinical workflows for providing care to mechanically ventilated patients in the ICU, a topic of immense public health importance, particularly in the wake of the COVID-19 pandemic. We deepen the understanding of the barriers to WUB adherence with particular attention to clinicians' information needs, coordination needs, and overall uncertainty. \textcolor{black}{Second, we identify novel opportunities for computational and artificial intelligence (AI) approaches to help clinicians effectively execute the patient liberation protocol} by emphasizing the `how' of protocol implementation, rather than simply the `what'. Based on these advances, we discuss design implications and present future research directions for better integrating protocols and guidelines into clinical practice beyond the context of mechanical ventilation in the ICU.

\section{Related Work}

\subsection{HCI Research on Intensive Care}
As an information-rich environment, the ICU has attracted the attention of HCI researchers over the past few decades. \textcolor{black}{Early ethnographic research investigating the clinical practice in the ICU noted the cyclic and temporal nature of activities: daily routines, such as patient rounds, helped clinicians plan and coordinate their tasks and overall patient care \cite{reddy2002finger}. Several studies highlighted the importance of presenting information in a temporal context, showing not only current activities, but also past and future activities that present the overall patient trajectory \cite{kusunoki2015designing, reddy2006temporality}. Other research studied workflows around specific tasks,} such as how physicians and nurses form volume therapy decisions \cite{kaltenhauser2020you} or \textcolor{black}{how nurses document patient notes and their workarounds to understand the limitations of electronic health records (EHR) \cite{collins2012workarounds}. Building on this understanding of clinical practice, researchers designed and evaluated the use of new technical systems to} support documentation and note-taking \cite{wilcox2010physician}, reduce alert fatigue and interruptions \cite{mastrianni2021designing, srinivas2016designing}, and support patient handovers \cite{segall2016operating}. In the same spirit, our work seeks to \textcolor{black}{understand the clinical workflows of nurses and RTs around WUB to inform the design of future technologies.}

\subsection{\textcolor{black}{Coordination in Clinical Settings}}

\textcolor{black}{Supporting collaboration and teamwork within complex healthcare settings has been a major research interest within the HCI and \textit{Computer-Supported Collaborative Work} (CSCW) literature \cite{fitzpatrick2013review}. Patient care is often delivered by multiple care providers that coordinate across time and space \cite{bossen2002parameters, bardram2000temporal}; however, coordination between care team members remains challenging \cite{amir2015care}. Prior work studied coordination in time- and safety-critical settings, such as surgical ward \cite{scupelli2010supporting, bardram2010plan, bardram2000temporal}, trauma resuscitation room \cite{sarcevic2011coordinating}, emergency department \cite{lee2012loosely}, and intensive care \cite{kaltenhauser2020you}. Researchers highlighted the role of artifacts and \textit{common information spaces} \cite{bossen2002parameters, hughes1997constructing}, such as paper schedules, electronic records, whiteboards and interactive displays, in facilitating coordination and collaborative sensemaking \cite{tang2007observational, xiao2007whiteboards, paul2010understanding, rambourg2018welcome, bjorn2011artefactual, bardram2006awaremedia}.}

\textcolor{black}{Early research viewed the shift from paper-based medical records to electronic records as an opportunity for technology to support collaboration between care team members by providing role-based, shared representations of care plans \cite{reddy2001coordinating, reddy2006temporality, reddy2002finger, reddy2003sociotechnical}. However, recent studies report that EHR systems lack collaborative affordances, making it difficult for clinicians to obtain the type of holistic understanding of the patient that is necessary to coordinate their activities \cite{karunakaran2012investigating, stevenson2010nurses, bardram2018collaborative}. Researchers note that current systems do not support team-based, shared care plans: team members often cannot access all of the patient information, and care goals are often set separately rather than as a team \cite{amir2015care}. As a result, many clinicians often view the EHR as a hindrance to collaboration, rather than an enabler \cite{chao2016impact, bates2010getting, o2010electronic}. Bardram and Houben \cite{bardram2018collaborative} highlight four design implications for medical records to facilitate collaboration: being portable across patient wards and the hospital, providing collocated access, providing a shared overview of medical data, and giving clinicians ways to maintain \textit{mutual awareness} – the ability to perceive each others’ activities and relate them to a joint context \cite{rittenbruch2009historical}.}
\textcolor{black}{We build on prior research on care coordination by investigating the use of EHR and other systems in the ICU for performing WUB – a non-emergency protocol that requires the sequential and coordinated actions of care providers.}

\subsection{Adherence to Clinical Guidelines}

The effective translation of clinically validated protocols and guidelines into daily clinical practice is considered one of the greatest challenges in evidence-based medicine \cite{stoneking2011sepsis, gary2021addressing}. A large body of healthcare literature has investigated the barriers to adherence in the context of ICU, including the use of guidelines for sepsis \cite{stoneking2011sepsis}, patient nutrition \cite{cahill2010understanding}, patient liberation from ventilators \cite{burns2012adherence}, and others (e.g., \cite{van2017development, bodi2005antibiotic, gurses2008systems}). These studies use a variety of methods, but typically approach the subject through a psychological lens, exploring barriers to behavior change on the part of the individual providers. Interestingly, the reported barriers using these methods are highly consistent across the literature, regardless of the specifics of each guideline. The barriers mainly include: lack of time and staffing, unfamiliarity with guidelines, disagreement with guidelines, and lack of coordination across clinical roles \cite{stoneking2011sepsis, gurses2008systems}. Researchers also report that guidelines often do not account for the complexity of the ICU environment, and can be perceived as a ``top down'' approach that does not consider the expertise or judgment of care providers \cite{gary2021addressing}.

Based on these challenges, medical researchers have proposed solutions to increase the uptake of clinical guidelines. These efforts typically fall into two broad approaches. One approach characterizes care providers' individual beliefs, attitudes, and lack of awareness as the root cause of noncompliance \cite{stoneking2011sepsis}. This strand of research proposes interventions such as formal education and hospital-wide training programs to align provider attitudes and beliefs toward evidence-based practice. While studies show that these interventions can help shift attitudes, researchers note that they are insufficient to change actual clinician behavior behavior \cite{blomkalns2007guideline, scott2007evolving}. \textcolor{black}{A second, less popular approach} characterizes adherence as a systems problem rather than solely a human problem \cite{gurses2008systems, lyon2019use}. This line of work suggests that noncompliance is a consequence of the interactions between care providers across time, locations, and system touchpoints along with physical and cultural components of an ICU. Gurses et al. frames current barriers as ambiguity around tasks, methods, and exceptions, and recommends interventions that reduce ambiguity (e.g., use of visual cues to indicate the status of patients with respect to a particular guideline) \cite{gurses2008systems}.

\textcolor{black}{Prior HCI research has explored how EHR systems influence protocol adherence \cite{pine2014institutional, katzenberg1996computer}, and developed systems to support the use of clinical guidelines in the form of digital checklists (e.g., trauma resuscitation \cite{kulp2019comparing, kulp2017exploring, mastrianni2023supporting}), treatment recommendations (e.g., providing clinical guidelines for pneumonia \cite{jones2019cds}, sepsis \cite{sivaraman2023ignore}, and cancer \cite{beauchemin2019clinical}), and guideline-specific dashboards (e.g., displaying a list of patients that qualify for a protocol \cite{anderson2019clinical}). While helpful in operationalizing broad best-practice guidelines, these approaches are often limited at the bedside as they do not account for patient-level variation \cite{klompas2020current}. Recent work highlighted the potential of AI technologies to combine clinical guidelines with patient-level variables to deliver more specific and personalized treatment recommendations \cite{sivaraman2023ignore}. We build on this line of research to understand information needs for WUB to discover opportunities for EHR-based and AI-based interventions that make it easier to consider and perform clinical guidelines.}

\subsection{\textcolor{black}{AI Systems in Healthcare and ICU}}

\textcolor{black}{A large body of research has explored data-driven and AI applications in healthcare, often in the form of clinical decision support systems (CDDS) (e.g., \cite{yang2019unremarkable, cai2019hello, sivaraman2023ignore}). Within the context of ICU, the majority of AI applications has focused on automating documentation related tasks to save time (e.g., transcribing ICU rounds \cite{king2023voice}) or providing diagnostic or prognostic insights to help with decision making (e.g., predicting the onset of conditions such as sepsis \cite{nemati2018interpretable, henry2022human}, tachycardia \cite{liu2021top} or hypotension \cite{yoon2020prediction}). Recently, with the availability of high-density EHR data on large numbers of mechanically ventilated ICU patients (e.g., MIMIC \cite{johnson2020mimic}), researchers created AI systems that predict if a patient will need a ventilator \cite{suresh2017clinical}, predict optimal ventilator settings for a patient \cite{peine2021development}, and predict the risk of patient extubation failure \cite{zhao2021development, torrini2021prediction}.}

\textcolor{black}{While proof-of-concept predictive models showcase initial feasibility, AI systems often fail when moving from research labs to clinical practice \cite{park2019identifying, yang2016investigating, hooper2012randomized, wilson2015automated}. HCI researchers point out that the clinical utility and actionability – the specific actions clinicians can take based on a prediction – of predictive models often remain unclear \cite{ghosh2023framing, tulk2022inclusion, yildirim2021technical, thieme2020machine}; and that seamless integration into current workflows is critical for clinician acceptance \cite{osman2021realizing, thieme2023designing, yang2016investigating}. In response, an increasing body of HCI literature has called for socio-technical, participatory approaches for understanding clinical workflows and engaging healthcare stakeholders early in the AI system development \cite{jacobs2021designing, cai2021onboarding, yang2016investigating, ghosh2023framing, yildirim2023investigating, robertson2020if, andersen2023introduction}. In the context of intensive care, a recent interview study explored \textit{what predictions would be useful} for ICU physicians and nurses \cite{eini2022tell}. Interestingly, clinicians expressed desires for predictions around patient trajectory and prioritization, mainly to reduce the high cognitive load caused by tracking the status of multiple highly dynamic patients rather than aiding decision making.} \textcolor{black}{Our research builds on this line of work by investigating current workflows for mechanically ventilated patient care with an eye for opportunities for clinically relevant AI prediction tasks to support the use of WUB in clinical practice.}

\section{Background on the Wake Up and Breathe Protocol}

ICUs are specialized hospital wards designed to care for patients at extremely high risk for death and disability, particularly patients experiencing acute organ dysfunction necessitating artificial life support. Mechanical ventilation provided via a tube inserted through the trachea (known as ``endotracheal intubation'') is among the most common forms of life-support provided in the ICU. Over 700,000 patients in the United States undergo mechanical ventilation in an ICU each year, typically for severe pneumonia or other types of acute respiratory failure, and overall mortality for these patients approaches 30\% \cite{wunsch2010epidemiology, hurley2023trends}. Recently, the COVID-19 pandemic caused an enormous influx of patients requiring mechanical ventilation in ICUs worldwide, highlighting the need for approaches to improve outcomes for this high-risk patient population \cite{bhatraju2020covid}.

Mechanical ventilators offer a life-saving intervention by both protecting a patient's airway and by taking over the work of breathing \cite{hickey2020mechanical}. This treatment helps manage a patient's oxygen level and provides support so their body can rest and heal. Ventilators manage the inflow of oxygen and outflow of carbon dioxide, allowing clinicians to control air volume, pressure, flow, oxygen levels, and respiratory rate. They can take over the entire work of breathing while patients are sedated to tolerate the endotracheal tube running through their throat or they can be set to assist, to react and support a patient's effort to breathe \cite{morandi2011sedation}. In the US, the use of a ventilator requires the collective actions of a team of specialists including the respiratory therapist (RT), who makes changes to the ventilator's modes and settings, the attending physician, who oversees the totality of a patient's care, and an ICU nurse, who monitors a patient's changing condition, adjusts sedation in real time \cite{willdenbest}.

Standards of care for ventilators have considerably changed over the last six decades; best practices suggest reducing time spent on ventilators, reducing sedation, and reducing breathing support \cite{gattinoni2020less}. Continued sedation can lead to delirium \cite{morandi2011sedation}, and remaining on a ventilator increases the chances of pneumonia \cite{hickey2020mechanical}. Air pressure and volume forced into the lungs can cause damage, and it can negatively impact heart function \cite{gattinoni2020less}. Too much oxygen can lead to toxicity in the blood \cite{gattinoni2020less}. Clinical teams work to get patients off the ventilator as quickly as possible to reduce the negative effects of their use. However, liberating patients from a ventilator too soon will frequently lead to reintubation, which is associated with greater morbidity and mortality \cite{robriquet2006predictors, richardson2020presenting, grasselli2020baseline}.

Medical researchers have created \textit{`the Wake Up and Breathe'} protocol that describes evidenced-based practices for patient liberation \cite{khan2014effectiveness}. First, an ICU nurse will perform a spontaneous awakening trial (SAT); they cut sedation and then assess if a patient can adjust to being awake. Next, an RT will perform a spontaneous breathing trial (SBT); they remove the ventilator's support and monitor if the patient can successfully breathe on their own. Based on the results of these two assessments, the nurse, RT, and physician will decide if a patient can be extubated (i.e., have the breathing tube removed). To make this decision, the team monitors the patient's physiological status during the SAT and SBT – patients who are able to breathe comfortably during the trials are eligible for extubation, while patients who show signs of respiratory distress during the trial should be returned to the full support of the ventilator. The protocol is performed every 24 hours for mechanically ventilated patients that meet the criteria to evaluate whether a patient is ready to be liberated from the ventilator \cite{willdenbest}.

The protocol requires careful coordination between the nurse who executes the SAT and the RT who executes the SBT. Cutting sedation, done by the nurse, should occur shortly but not immediately before the RT starts the breathing trial. Cutting the sedation too early may lead patients to become agitated and experience extreme discomfort from the breathing tube while waiting for the SBT to begin. However, too much sedation inhibits the drive to breathe and may cause the patient to fail an SBT. Typically, the assessment is initiated by the nurse in coordination with the RT. It does not require the direct involvement of physicians. \textcolor{black}{The protocol is ideally carried out before the patient rounds, so that during the rounds, the RT and the nurse report for each patient, allowing the care team to make the final determination about patient extubation. Once the protocol is delayed after rounds, clinicians tend to miss `the window of opportunity’ for WUB, which often postpones patient extubations until the following day, leading patients to remain longer on the ventilator \cite{balas2021evaluation}.}

The use of WUB reduces the average number of days a patient remains on a ventilator and decreases mortality \cite{khan2014effectiveness}. Unfortunately, the success of the protocol in controlled studies has been hard to reproduce in clinical practice. Often fewer than 50\% of eligible patients receive the SAT and SBT \cite{kher2013development, boehm2017perceptions}. Researchers have used surveys and interviews of ICU clinicians to probe barriers to protocol adherence \cite{kher2013development, costa2017identifying, gurses2008systems}. They note that nurses seem more reluctant to cut off sedation for patients receiving high doses \cite{kher2013development} as well as for patients who suffer from conditions like diarrhea, fatigue, confusion, or agitation \cite{costa2017identifying}. SAT and SBT are not performed when the workload is perceived as high or complex, or when the team is unfamiliar with the protocol \cite{costa2017identifying}. In addition, clinicians sometimes disagree with patient eligibility criteria \cite{costa2017identifying} or believe that SAT can cause the patient short-term harm \cite{miller2013diverse} and that a resting patient is healing \cite{costa2017identifying}. Finally, barriers also include a lack of coordination \cite{boehm2017perceptions, costa2017identifying} and a lack of supportive leadership \cite{costa2017identifying}.

Prior research explored how to overcome barriers by comparing the differences between ICUs that have better adherence to ICUs that have poorer adherence. They found the protocol is used more when teams can predict each other's actions \cite{boltey2019ability, costa2018icu}, and when the physician is involved in performing the daily breathing trials \cite{costa2018icu}. Additionally, discussion of sedation during daily patient rounds indicated a higher likelihood of the protocol being executed \cite{miller2013diverse}. A recent study showed that presenting a dedicated dashboard displaying eligible patients to nurses and RTs may shorten the duration of mechanical ventilation, although the effect sizes were modest and rates of SAT/SBT use was unchanged, calling into question the robustness of the findings \cite{anderson2019clinical}. Building on this line of research, this paper aims to understand the clinical practice and information needs around patient liberation in the ICU in order to inform more effective use of machine intelligence to enhance collaborative practice.

\section{Method}

We wanted to gain a deeper understanding of the clinical workflow around the liberation of mechanically ventilated patients. Specifically, we sought to understand how nurses and respiratory therapists decide which patients should receive SAT and SBT, and situations where they deviate from the protocol. We wanted to gain a nuanced understanding of previously identified WUB barriers and identify opportunities where technology might help.

\subsection{Study Design}

Building on prior ethnographic approaches in HCI research for healthcare \cite{reddy2006temporality, kaltenhauser2020you, yang2016investigating}, we chose to conduct a qualitative field study to capture clinical workflows and context with an eye for design opportunities to improve clinical practice. We conducted observations and semi-structured interviews across three sites in the east coast of the United States. All sites were medical ICUs (MICU), which are general facilities that treat adult patients with any number of serious medical conditions (as opposed to specialized ICUs such as surgical ICUs or pediatric ICUs). To strengthen the generalizability of our work, we studied both community and academic ICUs. The study was approved by our Institutional Review Board. Below we provide an overview of the sites:

\begin{figure*}
  \centering
  \includegraphics[width=2.1\columnwidth]{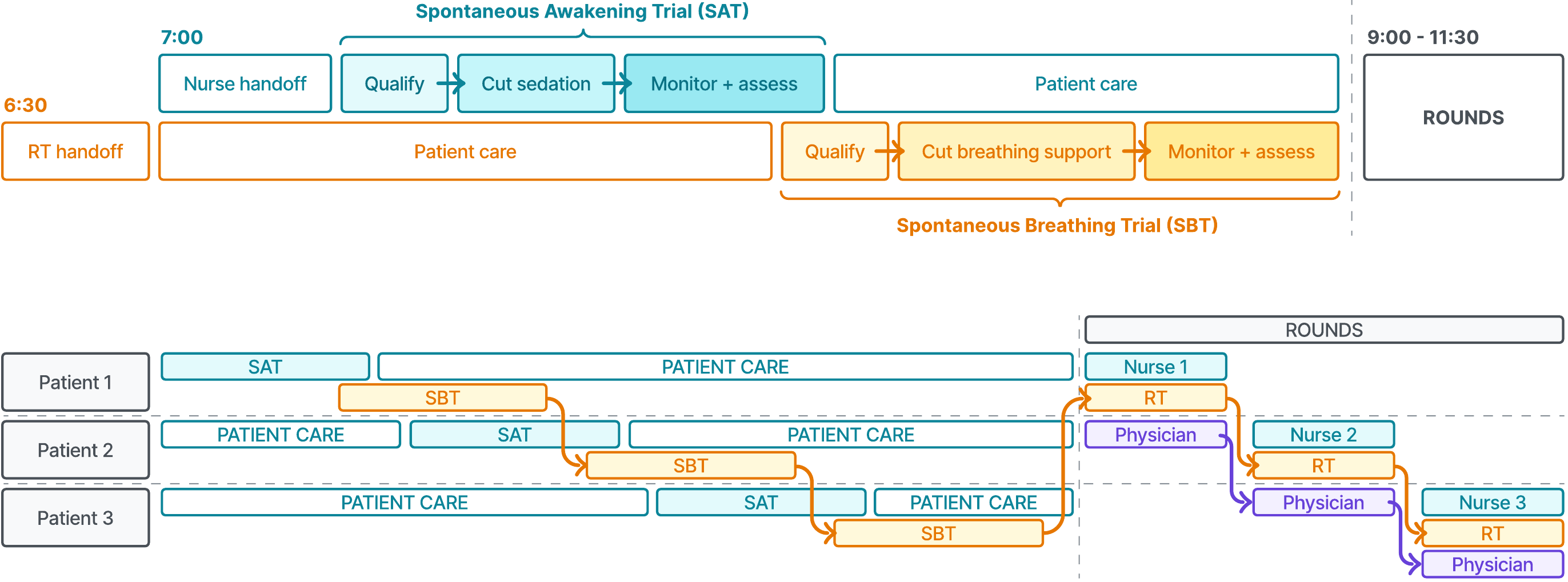}
  \caption{\label{fig:workflow}Ideal workflow for Spontaneous Awakening Trial (SAT) and Spontaneous Breathing Trial (SBT) for a single patient (top) and unit-level patients (bottom). \textcolor{teal}{Nurses (teal)}, \textcolor{orange}{RTs (orange)}, \textcolor{violet}{physicians and all other clinicians performing patient rounds (purple)} are shown as swim lanes. Notice that the RT ideally arrives at each patient's room shortly after the nurse caring for that patient has started the SAT.}
  \Description{A service blueprint documenting the workflow of nurses and RTs for SAT and SBT respectively. At the top, RT and nurse tasks for a single patient is shown with a focus on SAT/SBT timeline before patient rounds. At the bottom, a similar timeline is shown for multiple patients, where each patient is represented as a swim lane. RT can be seen moving between several patients to arrive timely after SAT started for a patient.}
\end{figure*}

\begin{itemize}
    \item \textbf{Site 1:} A community hospital with a 16-bed MICU. It has attending physicians who supervise residents, but it does not operate as a teaching hospital. The MICU includes step-down beds where patients receive a higher level of treatment than a typical inpatient ward but less intensive interventions and monitoring than a typical ICU.
    \item \textbf{Site 2:} An academic hospital attached to a cancer center. It has an 18-bed MICU and other specialized ICUs. The patients are seen by physicians and fellows who supervise residents and medical students. It primarily admits oncology patients, but it does treat other patients when needed.
    
    \item \textbf{Site 3:} An academic hospital with a 14-bed MICU. This site specializes in women's care, but it also provides care for all types of critically ill patients.
\end{itemize}

\textbf{Observations.} We observed clinical workflows over 17 sessions (47 hours in total). We observed care in patient rooms by reviewing the EHR and purposively selecting mechanically ventilated patients who are likely to be eligible for an SAT and SBT. We also followed the clinical team during rounds, and shadowed individual clinicians (i.e., attending physician, RT, nurse) in order to observe collaborative interactions from a variety of different perspectives. During observations, we recorded field notes, paying attention to start time and duration of clinical actions. We did not collect any personal identifiable information and did no recordings to protect patient privacy.

\begin{table}
  \caption{1:1 interview participants.}
  \label{tab:interview}
  \begin{tabular}{llllll}
    \toprule
    ID & Exp. & Site & ID & Exp. & Site\\
    \midrule
    Nurse 1 & 6 yrs & Site 3 & RT1 & 2 yrs & Site 2 \\
    Nurse 2 & 6 yrs & Site 3 & RT2 & 12 yrs & Site 2 \\
    Nurse 3 & 4 yrs & Site 1 & RT3 & 28 yrs & Site 1 \\
    Nurse 4 & 14 yrs & Site 1 & RT4 & 7 yrs & Site 2 \\
    Nurse 5 & 1.5 yrs & Site 1 & RT5 & 8 yrs & Site 2 \\
    Nurse 6 & 2 yrs & Site 1 & RT6 & 5 yrs & Site 1 \\
    Nurse 7 & 10 yrs & Site 2 & RT7 & 7 yrs & Site 1 \\
    Nurse 8 & 1 yrs & Site 1 &   &  &  \\
    Nurse 9 & 1 yrs & Site 1 &  &  &  \\
    Nurse 10 & 2.5 yrs & Site 2 &  &  &  \\
    \bottomrule
  \end{tabular}
\end{table}

\begin{table}
  \caption{Group interview participants.}
  \label{tab:group}
  \begin{tabular}{llll}
    \toprule
    Session & ID & Exp. & Site \\
    \midrule
    Session 1 & G1-Nurse 1 & 3 yrs & Site 1 \\
    Session 1 & G1-RT1 & 28 yrs & Site 1 \\
    Session 1 & G1-RT2 & 5 yrs & Site 3 \\
    Session 2 & G2-Nurse 2 & 5 yrs & Site 1 \\
    Session 2 & G2-Nurse 3 & 3 yrs & Site 2 \\
    Session 2 & G2-RT3 & 22 yrs & Site 2 \\
    Session 3 & G3-Nurse 4 & 1 yr & Site 1 \\
    Session 3 & G3-RT4 & 20 yrs & Site 2 \\
    Session 4 & G4-Nurse 5 & 1.5 yrs & Site 1 \\
    Session 4 & G4-RT5 & 5 yrs & Site 3 \\
    Session 4 & G4-RT6 & 19 yrs & Site 1 \\
    \bottomrule
  \end{tabular}
\end{table}

\textbf{1:1 Interviews.} We conducted one-on-one interviews with 10 nurses and 7 RTs. We used a semi-structured interview guide designed to elicit information about their role, practices and workflow around mechanically ventilated patients. We asked them to walk us through a recent patient where they performed an SAT and/or SBT in order to ground their responses in specific patient experiences. The open ended section of the guide probed for details around collaboration and coordination with clinical care team members; timing of care; and other related tasks they performed. We asked them about cases where the protocol might not be followed, current barriers and their needs for support. Interviews lasted between 45-60 minutes, and were audio recorded and transcribed verbatim. Participants were compensated \$100 for their time. Table~\ref{tab:interview} provides an overview of the participants from the one-on-one interviews.

\textbf{Group Interviews.} Following the one-on-one interviews, we conducted group interviews to validate the insights we gained with nurses and RTs together. In each session, we had at least one nurse and one RT participant, with a total of 11 participants (i.e., 6 RTs, 5 nurses) across four group sessions. Except for three participants, none of the participants had been involved in our prior interview study. Each session lasted 45-60 minutes. Participants were compensated \$125 for their time. Table~\ref{tab:group} presents an overview of the participants from the group interviews.

We recruited an initial set of participants through our collaborators at each hospital who shared our research study with their clinician colleagues. We then expanded this set through snowball sampling~\cite{Simkus_2023}, asking participants to share any contacts with the relevant clinical experience.

\subsection{Data Analysis}
We analyzed our notes from observations using affinity diagrams \cite{martin2012universal}. We created a service blueprint \cite{bitner2008service} of the clinical workflow for mechanically ventilated patient care. We analyzed the interview transcripts through thematic analysis, following the analysis process outlined in \cite{braun2006using}. First, all researchers familiarized themselves with the data. We then began generating initial codes and divided the transcripts between researchers. Each interview transcript was read and coded by at least two researchers. We frequently met to review, discuss and refine themes, and iteratively update the codes as we analyzed each interview. We generated 11 domain categories (e.g., SAT, SBT, timing of care, coordination) and 72 preliminary themes through several iterations and discussions, from which we constructed three main themes.

\section{Findings}
First, we provide a high-level overview of the clinical context for the patient liberation protocol in the ICU. Second, we provide evidence demonstrating the essential role of workflow uncertainty as a barrier to effective care delivery, capturing the various points of uncertainty that nurses and RTs face across the protocol steps, from assessing a patient's eligibility to receive a protocol, to performing the protocol and making care decisions (e.g., whether to liberate the patient or not). Finally, we detail how care providers interact with the EHR with a focus on their unmet information needs.

\subsection{Overview of the ICU Context}

During our observations, we did not see a single instance where WUB was executed as the protocol is described in the literature. SAT often did not happen or happened after an SBT started. Echoing prior literature, several participants shared that less experienced nurses might not be aware of the protocol or might feel intimidated to cut sedation: \textit{``They're afraid, it's very intimidating. [The patient is] biting at the tube … you're trying to make sure their restraints are tied.'' (Nurse 2)} However, the resistance to SAT seemed much less than the literature indicates. The lack of adherence did not stem from a lack of implementation: SAT and SBT were often performed, albeit with deviations.

Instead, what we observed was a tension between the desire to perform the protocol and the obligation to do a whole bunch of other tasks – paired with the notion that the other tasks are more urgent or easier as they do not require collaboration: \textit{`You want to do it, you just sometimes can't in reality … It's very busy, you're pulled in a lot of different directions.' (Nurse 7)} Below we present an overview of what a day looked like for a nurse and an RT with particular attention to the timeline for SAT and SBT.


\subsubsection{Team Composition and Tasks}

All ICU sites contained physicians, nurses, and RTs who were meant to collaborate when executing WUB. These different team members had varying schedules, assignments, and responsibilities. The ICU nurses worked three 12-hour shifts during a 7-day period. Each nurse has two patients, but they may have more if the ICU is short staffed. Nurses received patient assignments at the beginning of their shift and are not always assigned to the same patients day-to-day. The RT group was usually composed of 7-10 RTs – these care providers support the entire hospital and can be assigned to work on the ICU unit or elsewhere. When they are assigned to the ICU, typically one or two RTs are responsible for all patients in the unit (i.e., somewhere between 16 to 20 patients). The attending physicians worked 12-hour shifts over 7 consecutive days (though often staying longer). As a result, all care teams for individual patients, defined as the physician-nurse-RT triad responsible for a specific mechanically ventilated patient, were largely ad hoc in nature and would vary from patient-to-patient, day-to-day, and week-to-week.

The RT, nurse, and physician start the day shift respectively at 6.30am, 7am, and 8am (Figure~\ref{fig:workflow}). Shift change takes place with verbal handoffs within each discipline for about 30 minutes. Nurses hold a daily huddle, a 10-minute brief meeting before handoffs, where the charge nurse announces patient assignments. RTs and physicians typically do individual 10-minute rounds to quickly go through each patient before the handoff. At Site-1 and Site-2, physicians and nurses had a 30-minute meeting around 8.30am to go over patients \textit{``just to get to speed and be on the same page''}. Clinicians shared that typically 30\% of the patients in a unit are on mechanical ventilators, whereas 50\% would be considered high acuity (e.g., 7-8 patients in a 16-bed unit). While we did not conduct observations during the COVID-19 pandemic waves, clinicians noted that during the pandemic \textit{`the whole unit was on ventilators'}.

SAT or SBT was only one of the many tasks nurses and RTs perform on a daily basis. ICU nurses had a large number of tasks they had to complete before rounds, including head-to-toe patient assessment, delivering medications, tracking and documenting vitals, placing new orders, cleaning or mobilizing patients, and occasionally talking to family members. RTs provided care for patients who are on invasive ventilation (i.e., mechanical ventilation) or non-invasive ventilation (e.g., CPAP or BIPAP, nasal high flow cannulas). They spent at least 10 minutes going through each patient in the ICU, performing a variety of breathing related tasks before rounds started. All clinicians worked under time pressure with the need to react to unscheduled, high priority events. On several occasions we observed individual members of a care team being pulled away from a task to address a crisis, such as reviving a patient going into cardiac arrest. As an RT put it, the care team was constantly \textit{`putting out fires'}.

\subsubsection{Timeframe for SAT and SBT}
All ICUs conducted daily multidisciplinary rounds, where the clinical team assembled outside of a patient's room to discuss the case and collaborate over a plan of action for individual patients. Rounds usually started around 9am and took about two hours or longer, which meant that SAT and SBT should ideally be completed before 9am. SAT took about half an hour, as patients needed time to wake up from the sedatives. SBT took about an hour to observe if the patient could successfully breathe on their own. Overall, WUB took about 1-1.5 hours to perform, which meant that the care team should have a plan on which patients to SAT and SBT, and initiate the protocol as soon as they start shift. This was rarely the case; as aforementioned, morning time before the rounds was incredibly busy with lots of tasks, leaving little time for effective coordination between the time they arrive and the time rounds start.

Our interviews revealed that the protocol was often delayed or deprioritized, especially when care providers felt uncertainty about whether a patient should receive the protocol, and uncertainty around each other's schedule. Below, we present nuanced details on how these uncertainties impede the protocol adherence for nurses and RTs.

\subsection{Points of Uncertainty Surrounding the Execution of Wake Up and Breathe}
Our service blueprint (mapping the use of protocol in clinical practice) revealed three phases for WUB: (1) assessing a patient's eligibility to receive SAT/SBT, (2) performing SAT/SBT, (3) informing care decisions for patient liberation. In each phase, care providers faced uncertainties that made it challenging to adhere to the protocol. These uncertainties cascaded across care providers and led to unintentional noncompliance, given the ease in which providers could default to care practices for which there was less uncertainty. Below, we present the points of uncertainty for each phase and detail how these caused deviations.

\subsubsection{Uncertainty before performing SAT and SBT}

\textbf{Struggle to know which patients are eligible for SAT and SBT.} Our interviews revealed that both nurses and RTs face challenges in assessing a patient's eligibility to receive an SAT or SBT. Several nurses indicated that the protocol provided only a high level guidance and multiple opportunities to consider a patient non-eligible for vague and often subjective reasons (e.g., ``hemodynamic instability''): \textit{``We need a more concrete pathway as to when we're going to SAT and SBT somebody, and what are signs and symptoms to look for.'' (Nurse 7)} SAT eligibility criteria required nurses to evaluate both patient conditions (i.e. hypertension, open chest, upcoming surgery, etc.) and ventilator settings (i.e. concentration of oxygen, pressure levels). Interpreting ventilator settings created an additional barrier, especially for inexperienced nurses: \textit{``the nurse may not know how to interpret vent settings into weanable or extubatable.'' (RT4)}

RTs were dependent on information from nurses in order to assess a patient's SBT eligibility. They do not have the full picture of a patient because their lens is narrowly focused on the patient's respiratory status rather than the whole patient: \textit{``[The patient] could have chronically high or low blood pressure that I'm unaware of. [I might think] the heart rate is really high, maybe I shouldn't SBT them. [Nurse] can say, oh, that's normal for them.'' (RT1)} RTs also looked for information about a patient's daily care plan, and whether the patient is scheduled for a procedure that made them ineligible (e.g., dialysis, surgery). Some RTs shared that after they get their report, they try to listen in during the nurses' handoff to learn more about the patients. They expressed that it can be challenging to get this information from the EHR or by looking at the patient: \textit{``I might not be able to look at the IV (intravenous bag) pole and be able to tell what [sedatives] they're on. That's where the communication with my nurse comes in. (RT1)''}

Uncertainty combined with high workload under time pressure seemed to deprioritize WUB. Care providers were constantly \textit{``pulled in a lot of different directions'' (Nurse 7)}. This drove them to prioritize simple actions they know they must take (e.g., administering medications or charting in the medical record), and delay complex actions with ambiguity and uncertainty: \textit{``We usually wait until we complete rounds to ask about whether [physicians] want us to do a sedation interruption (SAT) or an SBT. Between nursing and respiratory, if there is a question, we feel safer addressing it in rounds and then doing those two tasks.'' (Nurse 1)} Both nurses and RTs wanted to know the physician's priority and goals for patient liberation to address uncertainty: \textit{``If you're really lucky, you'll get the physician that comes through and catches you before you start your round. You say, Hey listen, this is who I've got, what do you want us to do today? Like, what is your goal, so that it's my goal, so that we can get more people extubated?'' (G1-RT2)} Importantly, given the busy ICU environment, these otherwise well-intentioned ``delays'' often meant the patient would never receive an SAT and SBT at all, not because of any intentional decision but instead because of uncertainty around eligibility affecting overall care.

\subsubsection{Uncertainty when performing SAT and SBT}

\textbf{What does it mean to cut sedation?} Our interviews surfaced ambiguities in terms of whether sedation should be cut completely or reduced. Some nurses shared that it is common practice to have ``a touch of sedation'' during SBT for patient comfort; they did not consider this a deviation from the protocol: \textit{``Just because they're on some sedation doesn't necessarily mean we're suppressing their whole respiratory drive. We can just make them a little bit more comfortable so that they're able to be more compliant with the ventilator.'' (Nurse 6)} Others raised concerns, noting that patients were unlikely to pass SBT while on sedation: \textit{``I ask the nurse giving me report, why did [the patient] fail their SBT? They did the SBT on fentanyl. It's not like a true SBT in my eyes.'' (Nurse 7)} RTs often commented that they would try an SBT for qualifying patients, regardless of sedation: \textit{``You look at drips, look at the patient, look at the vials, look at the ventilator. And then you start a trial if you think they're able to do it, regardless of whether there is some sedation or not.'' (RT6)}

\textbf{What counts as sedation?} Related to cutting sedation, our conversations also revealed uncertainty around what counts as sedation. All nurses we interviewed shared that propofol and fentanyl were the most commonly used sedatives; however there were participants who did not consider these as sedatives for SAT: \textit{``I was always told sedation isn't a pain medication and fentanyl isn't sedation.'' (Nurse 8)} A few participants mentioned the use of additional sedatives that did not suppress the respiratory system, yet this did not seem to be common knowledge: \textit{``There's really only one that people can do a spontaneous breathing trial on ... Precedex, because that does not have any respiratory depression. It's a bit infrequent, but we do occasionally use it.'' (Nurse 6)} In this context, some providers might feel that they are successfully completing an SAT when they cut only some sedative medications, leading to well-intentioned non-adherence.

\textbf{Which patients should have priority?} Among our participants, there was no agreed upon approach for patient prioritization for SAT and SBT. During our fieldwork, we observed some RTs creating a prioritized patient list on a piece of paper. When asked whether and how they prioritize patients, every RT we interviewed shared their own way of prioritizing. Some started with the patients who were most likely to extubate: \textit{``I will generally go ahead to see who is on the lightest settings, then I begin to do a spontaneous breathing trial on them. One of my first go-to's was in the morning, who can I think about waking up and breathing to possibly get extubated by noon.'' (RT4)} Others simply followed the geographic location order, going patient to patient in order of where patients are located in the ICU: \textit{``Unless I'm told something specific about a patient, I'm starting at bed one and I'm working around room to room.'' (RT5)} This individual approach taken by different RTs makes it impossible for the nurses to infer when the RT might arrive at their patient and to use this inference to choose when to start an SAT meant to coordinate with the RT. 

\subsubsection{Uncertainty after performing SAT and SBT}

\textbf{How long is `good enough' before deciding if a patient should be liberated?} Another point of uncertainty was around when to end the protocol to inform patient liberation decisions. While the protocol suggests that one hour or less is sufficient to evaluate whether patients can breathe on their own, there did not seem to be a standard duration in clinical practice. RTs pointed out that the duration of SBTs rather depended on physicians: \textit{``Some physicians will be like, they pass an SBT. It's been an hour. They're doing well. Let's go ahead and extubate. Love that. Other physicians, it's been an hour. Okay. Let's give it another hour and another one. ... it's just kind of undefined and it feels like you're kicking the can down the road.'' (RT4)} Lack of agreement on protocol duration, combined with the delayed timing of care meant that nurses and RTs had to chase physicians to get them to make a decision: \textit{``You need to go bother the doctors: Hey, this guy's been on a wean for so long. Can we decide one way or the other?'' (Nurse 4)}

\subsubsection{Uncertainty in collaboration and coordination}

\textbf{When should SAT and SBT be performed?} Our interviews revealed that there was no standard timing of care for performing the protocol. While all participants shared that the protocol is supposed to be performed in the morning before rounds, they noted that it might be delayed until the afternoon if the unit is busy. Some nurses performed SAT first thing in the morning, whereas others preferred getting patient care tasks out of the way before cutting sedation: \textit{``Usually I'll get my other patient squared away and then I'll go in [the mechanically ventilated patient's] room. And the first thing I do is stop their sedation, then I go on to perform my assessment, give them their meds so that I'm in the room when they wake up.'' (Nurse 4)} However, mornings seemed to be the busiest time for nurses, and depending on the patients: \textit{``I don't feel like 9 am is a great time when you're trying to pass meds.'' (Nurse 7)} These statements suggest that the timing is driven more on convenience rather than intentional decision making.

\textbf{How to coordinate with the nurse/RT?} All participants emphasized that collaboration between the nurse and RT was essential for the protocol. The coordination for SAT/SBT was often informal: \textit{``Just the coordination, like, `Hey, I'm gonna cut my sedation'. [The RT will] come back when they see me in the room.'' (Nurse 2)} This seemed to work well for participants with more experience who have been working together for a long while: \textit{``Having a crew that works together on the regular, things happen quicker, patients recover faster. But when they're all brand new travel nurses and brand new people, everything is hard. We don't know how to communicate with one another.'' (RT5)} All nurses we interviewed seemed confident working with RTs, except for one nurse: \textit{``Treatment wise, some patients get breathing treatments and they do like ventilator care and stuff. I'm not too sure what [RT's] schedule is.'' (Nurse 5)}

Interestingly, several RTs reported that they do not wait for an SAT to perform SBT due to time constraints and challenges in coordination: \textit{``I need to get through those 16 beds, I want to get all my treatments done in a timely fashion. So that's the quickest way for me to get it done. I don't want to negate doing an SBT on a patient just because I can't work with the nurse at that point in time.'' (G1-RT2)} They shared that they will start an SBT during their first pass, and they will circle back for a second or third pass for SAT: \textit{``If their sedation is higher, and they're not able to SBT, then I can let the nurse know. That gives them time to finish their work, pass their meds and start the sedation vacation. So that I can start a second attempt.'' (G1-RT2)}

Some RTs expressed concerns with this practice: \textit{``We never skip the wean but [due to heavy workload] we might not do the start over multiple times or it might happen late in the day.'' (RT1)} Some nurses felt frustrated about the lack of coordination: \textit{``Sometimes [RTs] won't even tell the nurse, they'll just put 'em on an SBT and I'm like, well, they're still on sedation, you know?'' (Nurse 7)} Moreover, nurses did not want to cut sedation without knowing when the RT would come for an SBT for their patient: \textit{``Sometimes [nurses] don't want to cut patients' sedation because they just can't be there to watch the patient. They have to take their other patient to MRI and they're gonna be gone for two hours.'' (Nurse 4)} Uncertainty around each other's schedule seemed to be a major barrier to accurately performing SAT and SBT.

\vspace{-2mm}

\subsection{Interactions with EHR}
In this section, we detail how nurses and RTs interacted with the EHR and other IT systems with respect to WUB. We capture their pain points as well as their needs for support.

\textbf{Two IT systems are not better than one.} Details on WUB live on a different system than the EHR. Clinicians access both on the same computer, but it is not possible to see both at the same time. None of the computers we observed had two screens, which would make it possible to see guidelines and information about a specific patient at the same time.

The EHR system prompted nurses for SAT (i.e. \textit{``Have you done an SAT for this patient?''}). However, participants shared that the prompt is buried in an obscure location in the EHR rather than being integrated into current workflows: \textit{``It's under the iView, under like patient care, there's a section saying `started' sedation vacation … But I don't know if everyone knows about it.'' (Nurse 3)} Some nurses shared that they rarely used the iView interface: \textit{``I feel like sometimes it's not there and I've never really dug deep enough to see why it's not. But yeah, I have seen it fire a task for sedation interruption.'' (Nurse 4)} A nurse leader recalled a case where the hospital management received a complaint for noncompliance: \textit{``I think the manager got a complaint or whatever, saying nobody was documenting we did the sedation vacation. I don't think many people are doing it.'' (Nurse 1)}

\textbf{EHR does not assess which patients are eligible.} When asked how they assess patient eligibility for the protocol, all participants shared that they use the information passed during handoff along with their current knowledge of the patient, rather than using the EHR to identify eligible patients. While the eligibility criteria was often captured (e.g., whether the patient is paralyzed, whether the patient has an open chest, etc.), the system did not provide any support for assessing patient eligibility. Several participants pointed out that the EHR could do a better job in indicating contraindications to help assess which patients are eligible: \textit{``[If there was] a daily SBT SAT thing that fires, but then it fires if it's appropriate. It's smart enough in the background to say there's a contraindication present, seek out physician or something like that, that would be really cool.'' (G2-RT3)} This information would serve to reduce uncertainty by minimizing the cognitive burden necessary to assess the patient, so long as the data were transparent and interpretable. 

\textbf{The system only prompts for a task, but does not provide guidance on protocol.} While the system prompted for SAT and SBT in the form of an alert, inputting data into EHR did not seem to follow the structure of the pathway. Instead, clinicians relied on their memory to go through contraindications and carry out the steps. Nearly all participants brought up the need for an EHR that can support and reinforce the WUB clinical pathway (a flow chart that documents the specific implementation steps for SAT and SBT): \textit{``[When you click on the order set for the respiratory therapist] it brings up an archaic 2009 image of the pathway that hasn't been updated. [The system] is prompting a task for us, but having the pathway visible and easily accessible as [RTs are] getting ready to perform SBT would be extremely helpful.'' (G2-RT3)} Some RTs noted that training new personnel has been a challenge, as there has been a high turnover rate since the COVID-19 pandemic. Several nurses (Nurse 1, Nurse 2, Nurse 4, Nurse 7) also indicated that the SAT pathway was not clear, and suggested having step-by-step guidance in the EHR: \textit{``[An integrated module in the EHR] would help so that nurses could know the series of steps that should happen. If the person is agitated, then you go get the physician to assess the patient rather than just restarting their sedation.'' (Nurse 4)}

\textbf{SAT/SBT are separate ``checkbox'' tasks, EHR reinforces individual accountability rather than coordination.} Participants' reflections revealed that they prioritized their individual accountability in performing and documenting SAT/SBT over coordination. As a nurse put it: \textit{``When you have people that do their SAT, it's kind of like I'm just coming in and doing my assessment. They're not really concerned with coupling it with the SBT portion of it.'' (G3-Nurse 4)} Specifically the way these two tasks are designed in the EHR seemed to unintentionally separate nurses and RTs instead of helping them coordinate: \textit{``We're just chasing metrics, you know what I mean? Like, did I do it? And then once it's checked, okay, I'm done. Because everybody is busy. It's not that they're trying to get around it, it's just, that's their work.'' (G3-RT4)}

\textbf{Clinicians are penalized for cases with uncertainty, the system does not give them credit for postponed SAT/SBT.} The system interface for documenting SAT and SBT prompted clinicians to either record their execution of the protocol or document why the protocol was not performed (e.g., contraindication present, patient not eligible). The task was in the form of an alert that would prompt clinicians in the morning around 9am, and would require documentation within two hours. Missing the task window would be automatically captured as noncompliance. Both nurses and RTs shared that for patients with uncertainty, the protocol was often postponed for later in the day with the documentation option \textit{``MD order, contraindication''} – which led the system to disregard the succeeding clinician efforts: \textit{``It doesn't matter if I go back anytime throughout the day. Since I already documented that contraindication, I don't get credit for any more of an SBT that we would do that day [for that patient].'' (G2-RT3)}

Moreover, the EHR entry led to inaccurate data capture: \textit{``It's more of a data [limitation]. It would show that the RT didn't do an SBT on that patient that day. I would just make the assumption as her leader that that patient must not have been eligible, even though they were.'' (G2-RT3)} These statements suggest that the official records indicating that WUB was not adhered to may not be accurate. A nurse further elaborated that the system would not accept `continuous sedation interruption' (a patient who receives no sedation). The EHR would force them to document SAT as a discrete action: \textit{``You only get credit if you start and stop [sedation interruption]. You're not technically stopping [to increase sedation back] but the trial ended, patients are advanced through the trial.'' (G2-Nurse 3)} The system behavior made the unit look noncompliant, even though the protocol was performed accurately: \textit{``We are realizing we weren't getting credit for not hitting ``stopped''. So our [compliance] score was very low. So we're trying to re-educate our staff.'' (G2-Nurse 3)} Credit for compliance often meant serving the needs of an inflexible EHR more than providing care patients needed.

\textbf{No unit-level view of mechanically ventilated patients to monitor protocol status and extubation readiness.} RTs shared that they have access to a patient-level SBT dashboard showing the patient's SBT history, yet this was not designed as a shared view; none of the nurses had seen it before (except a nurse leader). Our group interviews revealed the need for a unit-level shared view: \textit{``I think a dashboard would be helpful, like, everybody's everywhere, all at once.'' (G3-Nurse 4)} Some participants suggested that having a shared view of SAT/SBT status of all patients would save time when getting a hold of physicians, as patient liberation decisions were mostly physician-driven: \textit{``I wish we had a clinical pathway where it was: we put them on the wean, now we get the gas, now the physician comes to see them now, and then we extubate. I feel like a lot of our time is spent going and telling people: Hey this is happening, what do you wanna do?'' (Nurse 4)}

\section{Discussion}
Despite providing best practices for improved patient outcomes, the implementation and uptake of clinical protocols and guidelines in clinical practice remains a persistent challenge \cite{gary2021addressing, stoneking2011sepsis}. Care providers face challenges in operationalizing the protocols from highly controlled clinical trials in messy, complex, and resource constrained day-to-day practice. Interestingly, the healthcare literature approaches this challenge largely as a problem with human behavior, often overlooking the opportunity for technology to change how people work. Recently, a few researchers suggested taking a more systemic approach to investigate not only clinicians' interactions with each other, but also their interactions with computing systems \cite{gurses2008systems}.

We approached this challenge using an HCI lens to investigate the lack of protocol adherence in the ICU, specifically in the context of WUB. Our field study revealed many barriers in how nurses and RTs interact with the EHR that negatively impacted protocol adherence. Current systems do not automate the assessment of patient eligibility, scaffold clinicians in making that assessment, nor support planning how a team of clinicians might efficiently assess all relevant patients within an ICU. We identified several opportunities for technology to better support the execution of WUB. At a higher level, these insights led us to reflect on what `protocol adherence' means, and on what the role and behavior of technology might be.

Below, we first discuss how our findings point to new design opportunities for 
technology to help clinicians effectively execute WUB. We then reframe `protocol non-adherence' and discuss opportunities for HCI research to explore better forms and behaviors of ICU technology. Finally, we reflect on open research questions for future investigation.

\subsection{\textcolor{black}{Implications for Designing Protocol-Based Care}}

A core goal of our study was to explore how healthcare computing systems might better support clinicians in executing WUB within the messiness of real world practice. Our study revealed that current EHR systems provide little support. The protocol seems to have been encoded in the EHR in an inflexible way that did not capture clinicians' mental models. It forces them to readjust their actions and go out of their way to receive electronic credit when they have complied. No information on collaboration gets captured nor do clinicians seem to be rewarded for working together across a set of patients.

Reflecting on these insights, we draw three implications to inform and inspire new healthcare computing systems in addressing adoption barriers to WUB in the ICU: automating/predicting a patient's eligibility for the protocol, guiding clinician actions through the steps of the protocol, and coordinating protocol actions across clinicians and the many patients they must support. We suspect most if not all of these design considerations will generalize to other clinical guidelines and protocols employed in the ICU, and that they will likely be valuable to clinical practice beyond the ICU.

\subsubsection{Support assessing patient eligibility for protocol}
Recent literature suggests that indicating which patients are eligible to receive SAT and SBT increases the adherence to the protocol \cite{anderson2019clinical}. Our findings echo this. Both nurses and RTs expressed challenges in assessing a patient's eligibility. They desired better support from technology. Current SAT and SBT tasks in EHR are not intelligent; the system does not take into account when a patient has contraindications, even though these are often captured in the patient record. A better system might pre-qualify patients to indicate if they are eligible to receive a protocol. It might make delaying SAT a more explicit task for nurses and RTs, so that they seek physician input to address their uncertainty. \textcolor{black}{Notably, our study raises issues around data trustworthiness; patients with contraindications might have been simply documented as ineligible due to delayed decision making. Future work should investigate better data collection and data fusion (e.g., patient records, data from mechanical ventilators, etc.) to assess the feasibility of pre-qualifying patients for WUB.}

\textcolor{black}{We also see opportunities for AI-based systems to support clinicians in assessing patient eligibility for clinical guidelines. Recent research demonstrates the use of large language models (LLMs) to extract patient information from team discussions in ICU rounds in order to suggest evidence-based practices and guidelines \cite{king2023voice}. Future work can explore summarizing patient trajectory with regards to specific guidelines (e.g., what was discussed about patient extubation in the past week during rounds?). Additionally, as foundation models open up new avenues in extracting patient variables from EHR \cite{adamson2023approach, nori2023capabilities}, information from multiple sources (rounds, EHR, unstructured notes from nursing and RT documentation) can be combined to present patient-specific, contextual clinical guidelines. Indicating patient eligibility for a protocol is likely to increase the uptake of guidelines while reducing clinicians’ uncertainty and anxiety around whether they will receive credit for their work.}

\subsubsection{\textcolor{black}{Support clinicians in adapting clinical guidelines}}
\textcolor{black}{Prior work highlighted that clinical guidelines should account for the complexity of the ICU environment for increased adoption in clinical practice \cite{gary2021addressing}. Our work raises additional design considerations around the granularity and adaptability of clinical guidelines. Our findings point out that guidelines can remain too high-level, leading to ambiguities in successful execution. Building on the large body of literature on medical checklists \cite{mastrianni2023supporting, kulp2019comparing}, we see opportunities creating team-based, shared WUB plans that allow clinicians to set goals as a team – a critical aspect for improved care coordination \cite{amir2015care}. Future research should explore how clinical guidelines can be designed in a way that allows clinicians to customize, adapt, and add more granularity onto existing guidelines to operationalize them in their specific clinical context (e.g., offering clarity on what sedatives suppress the respiratory system, highlighting the ones a specific patient is receiving, and nudging nurses to cut sedation completely rather than reducing it). Central to this point is that computer systems could motivate clinicians to effectively perform the protocol, rather than penalizing them.}

\subsubsection{Support unit-level coordination of protocol actions between clinicians}
\textcolor{black}{HCI and CSCW literature has highlighted the importance of \textit{common information spaces} and \textit{mutual awareness} to help clinicians coordinate care in complex healthcare settings. Our study shows that role-based views can lead to information silos when they are not presented as team-based, shared plans. A better system could support unit-level coordination of SAT and SBT actions between nurses and RTs, rather than separating these as individual responsibilities. We see an opportunity for HCI research to explore interaction designs that present the work for multiple care providers across multiple patients. For instance, a dedicated SAT/SBT dashboard could present a shared view of all ventilated patients to nurses, RTs, and physicians to indicate patient status (e.g., SAT/SBT: ready, in process, completed, not eligible) as well as patient trajectory (e.g., days on the ventilator, prior SAT/SBT performance, prior unsuccessful extubations, etc). Anderson et al.’s study presents a great example in this direction \cite{anderson2019clinical}.}

\textcolor{black}{We also see opportunities for data and AI-based approaches for prioritizing patients in busy wards – a major pain point where clinicians think AI can help \cite{eini2022tell}. Prior research has mostly focused on prediction tasks around patient extubation for clinical decision support (e.g., whether a patient should be extubated based on the risk of extubation failure \cite{zhao2021development, torrini2021prediction}). This is a high-risk decision that brings up likely challenges around clinician acceptance \cite{yang2016investigating}. Instead, researchers can focus on predictions around WUB to support care coordination and prioritization. For instance, predictions of the likelihood of a patient to receive SAT/SBT (e.g., highly likely versus less likely) can help nurses and RTs coordinate WUB efforts for high priority patients (e.g., highlighting top three patients that should receive the protocol). Additionally, identifying patients with high uncertainty (e.g., patients with dynamic conditions who might become eligible later in the day) can prompt nurses and RTs to resolve uncertainties for patients who might otherwise not receive the protocol. Exploring whether and how this information can be best incorporated into clinical workflows marks a clear direction for future research.}

\subsection{Reframing Non-Adherence}
A large body of healthcare research has investigated the barriers to the adoption of clinical guidelines in practice \cite{blackwood2014protocolized, costa2017identifying, stoneking2011sepsis, gary2021addressing}. The majority of this literature approaches this challenge using the lens of \textit{compliance}: clinicians are `non-compliant' when they do not adhere to guidelines. This view has an inherent assumption that clinicians are either negligent as they are not aware of the existence of a protocol or they fail to adequately perform them, or they are non-compliant as they ignore or actively choose not to follow guidelines.

Our findings suggest that in the case for WUB, this characterization of clinician negligence does not capture the real story. Current literature suggests clinicians' lack of adherence stems from their reluctance or resistance to the protocol. However, what we observed and heard during the interviews indicates something closer to \textit{procrastination}: clinicians are overwhelmed with way more tasks than they can complete. When they experience uncertainty around a patient's eligibility for a particularly unpleasant task –such as cutting a patient's sedation or breathing support as the patient anxiously struggles to remove their tube– and uncertainty around the care team's priorities and schedule, they default to prioritizing tasks with high certainty. They do not explicitly choose not to perform a required task. Instead, they choose to prioritize tasks where they have certainty, and this unintentionally causes the uncertain tasks to remain undone. Similarly, clinicians' reflections on how they prioritize patients for WUB indicate that patients with high certainty for extubation seem most likely to receive the protocol. They prioritize and adhere to the cases where they think the protocol is most important. Their behavior feels a lot like triage, the allocation of resources where they will have the greatest impact, a normal part of clinical practice.

These insights provide the opportunity to reframe what `protocol adherence' means in the clinical setting. A new way to see this is as supporting how several clinicians from different roles work across different patients to execute an interrelated set of actions. There's an opportunity for ICU level coordination. Current EHR systems seem to unintentionally hinder this work by separating individual clinician's actions. Most healthcare IT systems privilege a patient-level treatment view for a single clinician performing care at a single time and place \cite{yang2016investigating}. A better healthcare system could be designed with a focus on the collective delivery of care, where the patient-level EHR is a component of a larger system that has a better knowledge of activities across all patients and clinicians \cite{glaser2020s}. For example, many shipping companies provide a customer-centric view, showing people where their package is, as well as a logistics view for executing an efficient delivery of packages across many packages, customers, and delivery drivers. We suspect this alternative, \textcolor{black}{service-oriented view \cite{roto2021overlaps, yildirim2022experienced}} will inspire new possibilities for computational systems that can support a coordinated set of activities, more efficiently and fairly allocating clinical efforts and skills across many patients to maximize outcomes for all. Recent work (e.g., \cite{hopkins2023designing, zajkac2023ground, noortman2022breaking, bogers2018situated}) exploring how to design datasets with domain stakeholders provide great starting points for this strand of research. We see an opportunity for HCI research to envision new forms and behaviors for technical systems and datasets that help clinicians effectively and collectively perform their day-to-day work.

\subsection{Open Research Questions}
Below we detail three research questions our study raised that merit further investigation.

\subsubsection{\textcolor{black}{How to design an AI system that coordinates interactions between care team members for increased use of protocols?}}
\textcolor{black}{While our study reveals opportunities for designing AI systems that support coordination between nurses and RTs, there are open questions from an interaction design standpoint. For instance, a ranked list of mechanically ventilated patients (e.g., based on eligibility and predicted WUB performance) can help RTs follow a specific patient order, so that nurses might better be able to predict RTs’ actions. However, questions remain regarding whether RTs and nurses will accept and use this type of AI system. We anticipate that the human-AI interaction design will be critical for communicating the criteria for patient prioritization in order to establish trust without hindering agency \cite{khairat2018reasons, yang2016investigating} (e.g., high priority patient due to successful SAT/SBT in the past few days). On the other hand, prior HCI work on algorithmic management systems note that such systems can easily become tyrannical tools (e.g., worker surveillance issues raised by the use of an algorithmic room assignment system in the hospitality industry \cite{spektor2023designing}). Future research should follow participatory approaches to co-design future technology with clinicians in ways that support them in coordinating while at the same time protecting their autonomy.}

\subsubsection{\textcolor{black}{How to situate protocol coordination systems in clinical workflows?}}
\textcolor{black}{In addition to interaction design questions around presenting patient eligibility and priority, there are open research questions around situating this type of systems in critical care workflows. When, where, and in what form should WUB information be presented to nurses, RTs, and physicians? For example, an SAT/SBT dashboard with draft patient priority can be reviewed by the night shift physician fellow for them to mark high priority patients before the day shift handover. Moreover, such a dashboard can be placed in patient rooms or can be situated in the ICU corridor as a mobile, interactive display for use during rounds. Future work should explore specific interaction designs and how clinicians might interact with these in-situ.}

\subsubsection{Does using AI to recommend or draft coordination plans offer an opportunity for lower-risk AI in healthcare? Would deploying these types of systems make clinicians more open to collaborating with AI in clinical practice?}
Recent research exploring AI innovation in healthcare proposed patient triage, workload management, documentation, and resource planning as lower-risk application areas that might be better starting places for integrating AI in healthcare compared to high-risk, critical applications such as clinical decision support \cite{yildirim2023creating, bossen2023batman}. \textcolor{black}{This provides a counterintuitive perspective, given that AI systems are increasingly more capable with the rise of multimodal foundation models \cite{lee2023benefits, nori2023capabilities}. Our findings bring up a similar point.} From a responsible AI perspective \cite{thieme2023foundation, zajkac2023clinician}, supporting clinicians in the execution of clinical guidelines seems to be less risky, both in terms of clinician acceptance and regulatory challenges. We encourage HCI and AI researchers to further investigate how computational systems support the translation of clinical guidelines into practice.

\section{Limitations}
Our study has three limitations. First, our fieldwork focused on nurses and RTs as the key clinician roles involved in the execution of WUB. However, there are many other roles involved in the broader context of patient liberation from mechanical ventilation, including physicians, patients, and patient's family members whose perspectives are not covered in this work. Future work should recruit a broader set of participants to understand their involvement. Second, our qualitative approach mostly focused on capturing the current state with little insight into clinicians' desired futures. Future research should take participatory, generative approaches to explore how clinicians envision computational systems to better support the incorporation of protocols in day-to-day practice. Finally, our study had a North American focus with its involvement of RT as a distinct clinical role. We acknowledge that ICU settings and practices may differ based on hospitals and geographical locations. Future research should investigate other settings to provide insights into how these findings may differ or generalize to other ICUs.

\section{Conclusion}
In this paper, we have presented a field study to understand why clinicians deviate from the clinical practice guidelines for liberating patients from mechanical ventilation in the ICU. We expanded prior literature by capturing the clinical workflows of nurses and RTs for mechanically ventilated patients, and identifying opportunities where technology might support clinicians in executing the patient liberation protocol. These findings suggest an alternative perspective for clinical protocol adherence, proposing that computational systems should work in service of clinicians to make it easier to consider and perform protocols, rather than characterizing clinicians as noncompliant. We strongly encourage HCI researchers to create socio-technical systems that can support the adoption and execution of clinical practice guidelines in complex, real world healthcare settings within and beyond the ICU.

\begin{acks}

We thank the participants in this work for their time and valuable input. This material is based upon work supported by the National Science Foundation under Grant No. (2007501) and work supported by the National Institutes of Health (R35HL144804). The first author was also supported by the Center for Machine Learning and Health (CMLH) Translational Fellowships in Digital Health. Any opinions, findings, and conclusions or recommendations expressed in this material are those of the authors and do not necessarily reflect the views of the National Science Foundation or the National Institutes of Health.
\end{acks}

\bibliographystyle{ACM-Reference-Format}
\bibliography{icu-fieldwork}










\end{document}